\PassOptionsToPackage{hyphens}{url}
\documentclass[sigplan,screen]{acmart}
\pdfoutput=1

\usepackage[pdf]{graphviz}

\usepackage{fancyvrb}
\usepackage{todonotes}

\newcommand{\FILE}[1]{}

\setcopyright{none}
\renewcommand\footnotetextcopyrightpermission[1]{}

\definecolor{friendlybg}{HTML}{f0f0f0}
\definecolor{lightgray}{rgb}{.9,.9,.9}
\definecolor{darkgray}{rgb}{.4,.4,.4}
\definecolor{purple}{rgb}{0.65, 0.12, 0.82}

\acmConference[von Laszewski, et al.]{University of Virginia}{Oct. 25, 2022}{Charlottesville, VA}

\begin{document}

\title{Hybrid Reusable Computational Analytics Workflow Management with Cloudmesh}

\author{Gregor von Laszewski}
\email{laszewski@gmail.com}
\orcid{0000-0001-9558-179X}
\authornote{Corresponding author}
\affiliation{%
  \institution{University of Virginia}
  \streetaddress{Biocomplexity Institute\\
Town Center Four\\
994 Research Park Boulevard}
  \city{Charlottesville}
  \state{VA}
  \postcode{22911}
  \country{USA}
}

\author{J.P. Fleischer}
\email{jacquespfleischer@gmail.com}
\orcid{0000-0002-1102-1910}
\affiliation{%
  \institution{University of Virginia}
  \streetaddress{Biocomplexity Institute\\
Town Center Four\\
994 Research Park Boulevard}
  \city{Charlottesville}
  \state{VA}
  \postcode{22911}
  \country{USA}
}

\author{Geoffrey C. Fox}
\email{gcfexchange@gmail.com}
\orcid{0000-0003-1017-1391}
\affiliation{%
  \institution{University of Virginia}
  \streetaddress{Biocomplexity Institute\\
Town Center Four\\
994 Research Park Boulevard}
  \city{Charlottesville}
  \state{VA}
  \postcode{22911}
  \country{USA}
}

\renewcommand{\shortauthors}{von Laszewski, et al.}

\begin{abstract}

 \FILE{abstract.tex}

 In this paper, we summarize our effort to create and utilize a {\em
 simple} framework to coordinate computational analytics tasks with
 the help of a workflow system. Our design is based on a minimalistic
 approach while at the same time allowing to access computational
 resources offered through the owner's computer, HPC computing
 centers, cloud resources, and distributed systems in general. The
 access to this framework includes a simple GUI for monitoring and
 managing the workflow, a REST service, a command line interface, as
 well as a Python interface. The resulting framework was developed for
 several examples targeting benchmarks of AI applications on hybrid
 compute resources and as an educational tool for teaching scientists
 and students sophisticated concepts to execute computations on
 resources ranging from a single computer to many thousands of
 computers as part of on-premise and cloud infrastructure. We
 demonstrate the usefulness of the tool on a number of examples. The
 code is available as an open-source project in GitHub and is based on
 an easy-to-enhance tool called cloudmesh.
 
\end{abstract}




\settopmatter{printfolios=true}

\maketitle

\FILE{cc.tex}

\section{Introduction}

In this section we provide an introduction to our work while
moving forward to motivate a 
Hybrid Reusable Computational Analytics Workflow
Management Framework.

\subsection{Reusable Computational Analytics}

{\em Reusable computational analytics} (RCA) focuses on the creation
of reusable programs, patterns, and services to conduct analytics
tasks that are part of the scientific discovery process. RCA service
need varies widely and may include multi-scale hardware resources as
well as multi-scale scientific applications. To utilize such services
and their resources in a reusable way, we need to have a mechanism to
express them in an easy fashion that goes beyond just the definition
in one programming language or framework, but allows the integration
into many different programming languages and frameworks so that
services that may be designed in one framework or language may be
reusable in others.

\subsection{Reusable Multi-scale Algorithms}

In current scientific problems we encounter a rich set of applications
that leverage a number of sophisticated methods that may require
adaptations on multiple scales. The scales are influenced by their
Domain size, accuracy and time requirement to solve them in a
sufficient manner. It is of advantage to provide reusable components
that can be controlled by parameters to simplify reuse.

\subsection{Hybrid Cloud and Compute Resources and Serices}

As we deal with multi-scale algorithms, not every analytics task needs
to be conducted on a High Performance Computer (HPC). This is
especially the case with the advent of desktop GPUs, which authors
have termed in past {\em desktop supercomputing}. Also the
availability of cloud computers and hyper-scale data centers play a
significant role in today's analytics processes. This not only
includes the use compute resources, but also services that are these
days offered by cloud service providers. A well-known example for this
is natural language processing.

\subsection{Reusable and Adaptable HPC and Cloud Service Workflows}

High-performance computing (HPC) has been, for decades, a very important tool
for science. Scientific tasks can leverage the processing power of
a supercomputer so they can run at previously unobtainable high speeds
or utilize specialized hardware for acceleration that otherwise are not
available to the user. HPC can be used for analytic programs that
leverage machine learning applied to large data sets to, for example,
predict future values or to model current states. For such
high-complexity projects, there are often multiple complex programs that
may be running repeatedly in either competition or cooperation.
Leveraging computational GPUs, for instance, leads to several times higher
performance when applied to deep learning algorithms. With such
projects, program execution is submitted as a job to a typically remote
HPC center, where time is billed as node hours. Such projects must have
a service that lets the user manage and execute without supervision. We
have created a service that lets the user run jobs across multiple
platforms in a dynamic queue with visualization and data storage.

Similar aspects are available for cloud services that abstract the
infrastructure needs and focus on the availability of services that
can be integrated in concert with HPC, as well as the users local
resources (for example a PC).

\FILE{requirements.tex}

\section{Requirements}

For the design of the framework, we have the following requirements.

\begin{description}

\item[Simplicity.] The design must be simple and the code base must be
small.  This helps the future maintenance of the code, but also allows
the code to become part of educational opportunities such as in a
Research Experience for Undergraduates (REU), capstone project, and
class projects. The challenge here is that many other systems are too
big to be used in introductory undergraduate and masters activities.
However, the framework should be capable enough to also support
research projects.

\item[Workflow specification.] The specification of the workflow must
be simple.  From the past we have learned that the introduction of
programming control components such as loops, and conditions, in
addition to DAGs, is essential to provide maximum flexibility.

\item[Workflow Monitoring.] The workflow framework must be able to
monitor the progress of individual jobs as well as the overall
progress of the workflow.

\item[Workflow Interfaces.] To specify and interface with the
workflow, we must provide several interface layers. This includes
specification through a YAML file, interfacing through REST calls,
interfacing through a Python API, and interfacing with a very simple
GUI component.  For the workflow monitoring with the GUI, we must be
able to easily define custom text to reflect user designable
monitoring labels.

\item[Language Independence.] As we want to make the framework
integrable in other frameworks, we need a simple mechanism to provide
either API or interface portability. To keep the code base small, a
REST API seems very well-suited.

\item[OpenAPI.] To further strengthen usability, the framework must
have an OpenAPI interface. This allows the integration via 3rd party
tools into other frameworks and languages, if desired.

\item[Hybrid multicloud Providers.] The service must be able to be
deployable on an on-premise local computer or various cloud providers.

\item[Generalized Job Interface.] The framework must be able to
interface with a wide variety of computation services. This includes
the support of ssh, batch queues such as Slurm and LSF, and local
compute resources including shell scripts, as well as support for WSL.

\item[Support of various Operating Systems.] The framework must be
runnable on various client operating systems including Linux, macOS,
Windows.

\item[State Management.] The state management must be recoverable and
the client must be able to completely recover from a network failure.
As such, the state will be queried on demand. This allows one to
deploy the framework on a laptop, start the workflow, close or
shutdown the laptop, and at a later time open the laptop while the
workflow framework can be refreshed with the latest state.

\end{description}

\FILE{related-research.tex}

\section{Related Research}

Many different workflow systems have been developed in the past. It is
out of the scope of this document to present a complete overview of
all the different systems. Instead, we compare some selected features
of the systems and identify features that are important for us. The
following features are important.

\begin{description}

\item[Batch paralelism.] Long-running jobs on HPC systems are coordinated through
Batch services.

\item[Task paralleleism.] Tasks are a logical unit of work that is
executed by a resource, service, or component. Often trask paralelism
is used in distributed resource frameworks.  

\item[Resource Reservation.] In some cases, access to batch queues on
HPC systems take a long time. In the case of many different tasks, it
is sometimes useful to research a number of batch nodes and run many
short-running programs on them if the need arises. This, however, can
often be replaced just with properly coordinated workflows using a
batch system. In fact, frameworks using such reservations internally
implement them using the batch system.

\item[Computational Grids.] Grids provide service-level access to
distributed HPC computing resources. However, Grids (popular a decade
ago) are no longer predominantly deployed and the focus has shifted to
Cloud computing.

\item[Cloud.] Presently, computing resources are also available in
the cloud as HPC, batch, and compute services, including specialized
SaaS offerings that allow to integrate analytics functions into
workflows.

\item[REST.] The predominant specification for cloud services uses the
REST framework relying on a stateless model. This contrasts with the
WSRF model that uses stateful services.

\item[Specification.] Some work has also focused on the specification
of the workflows.

\item[GUI.] Some frameworks provide extensive GUIs.

\end{description}

In Table~\ref{tab:workflow-table}, we provide a number of examples for
the various features we found in workflow systems.

\begin{table*}[htb]

\caption{Example of workflow frameworks with selected features.}
\label{tab:workflow-table}

\resizebox{1.0\textwidth}{!}{
{\footnotesize
\begin{tabular}{|p{3cm}|p{3cm}|p{10cm}|}
\hline
{\bf Name} & {\bf Selected Features} & {\bf Description} \\
\hline
\hline

LSF \cite{www-lsf} & Batch, HPC & Batch Queue Manager. \\
\hline

SLURM \cite{www-slurm} & Batch, HPC & Batch queue workflow manager. \\
\hline

Cloudmesh  \cite{www-cloudmesh-manual} & HPC, Cloud, on-Premise, Task & Suite
of Python code for cloud computing. \\
\hline

Airflow \cite{www-airflow} & HPC, Cloud, Remote Tasks & ``Airflow is a platform created by the community to programmatically author, schedule, and monitor workflows.'' \\
\hline

Loosely coupled Metacomputer \cite{las-1996-thesis,las-1999-loosely} &
                                                                       HPC, on-Premise
             & Early work on workflow management system connecting multiple supercomputers. \\
\hline

Karajan CoGkit  \cite{las07-workflow} & Service, Language, HPC, on-Premise & Sophisticated Workflow management system with Dag and loops for Grid computing and loosely coupled compute resources through services. \\
\hline

Snakemake  \cite{www-snakemake} & Language & ``The Snakemake workflow management system is a tool to create reproducible and scalable data analyses. Workflows are described via a human-readable, Python-based language. They can be seamlessly scaled to server, cluster, grid, and cloud environments, without the need to modify the workflow definition. Finally, Snakemake workflows can entail a description of the required software, which will be automatically deployed to any execution environment.'' \\
\hline

Gridant  \cite{las-2004-gridant} & Language & 
Client-based workflow management toolkit to orchestrate complex workflows on the fly without substantial help from the service providers. It integrates on premise and Grid resources. \\
\hline

Keppler  \cite{www-kepler} & Service, GUI & ``Kepler is designed to help scientists, analysts, and computer programmers create, execute, and share models and analyses across a broad range of scientific and engineering disciplines.'' \\
\hline

Pegasus  \cite{www-pegasus} & DAG, Service, HPC, Cloud &  A DAG-based workflow management tool for scientific workflow.  \\
\hline

Swift \cite{las-2007-swift} & Language, Resource Reservation & A language for distributed parallel scripting.\\
\hline

Parsl \cite{www-parsl} & Language, Resource Reservation & A language for distributed parallel scripting. \\
\hline

Radical Pilot \cite{arxiv-radical-pilot} & Resource Reservation & Pilot system that enables scalable workflows \\ 
\hline

Gcloud workflow \cite{www-gcloud} & Cloud & Google Cloud documentation for creating a workflow with the command line interface. \\
\hline

Azure REST \cite{www-azure-rest} & Cloud & Microsoft Azure REST API documentation to interface with Azure Logic Apps. \\
\hline

GitHub REST Cancel a Workflow Run \cite{www-github-rest-cancel}
& GitHub Actions
& GitHub REST API documentation to cancel a GitHub Workflow. \\
\hline


Azure Enterprise Workflow \cite{www-azure-enterprise-workflow} & GUI,
Workflow & Enterprise Workflows with Azure Logic Applications \\
\hline

WSDL \cite{www-wsdl} & Specification & Microsoft Azure schema reference for the Workflow Definition Language in Azure Logic Apps. \\
  \hline

  WSRF \cite{www-wsrf} & Specification & OASIS Web Services Resource Framework\\

AWS Workflow \cite{www-aws-workflow} &  & Amazon Web Services documentation definition of a workflow.\\
\hline

AWS SWF \cite{www-aws-swf} & Task & Amazon Web Services documentation for managing simple workflows with tasks. \\
\hline

AWS Step Functions \cite{www-aws-stepfunctions} & GUI, Task & Amazon Web Services documentation for visual workflow service. \\
\hline

AWS HPC  \cite{www-aws-batch-workflow} & Cloud, HPC & Amazon Web Services documentation for setting workflow dependencies using batch processing in HPC on the cloud. \\
\hline

Azure Batch \cite{www-azure-batch} & Cloud, Batch & Microsoft Azure documentation for running batch service workflows. \\
  \hline

 Business Automation with REST \cite{www-business-rest-ibm} & Business
                                        Process& IBM REST API documentation for its Business Automation Workflows. \\
\hline



Infogram \cite{las-02-infogram} & Service & A Peer-to-Peer Information and Job Submission Service.\\
\hline
\end{tabular}
}
}
\end{table*}

\FILE{cc-design.tex}

\section{Design}

\begin{figure}[htb]
{\centering
\includegraphics[width=1.0\columnwidth]{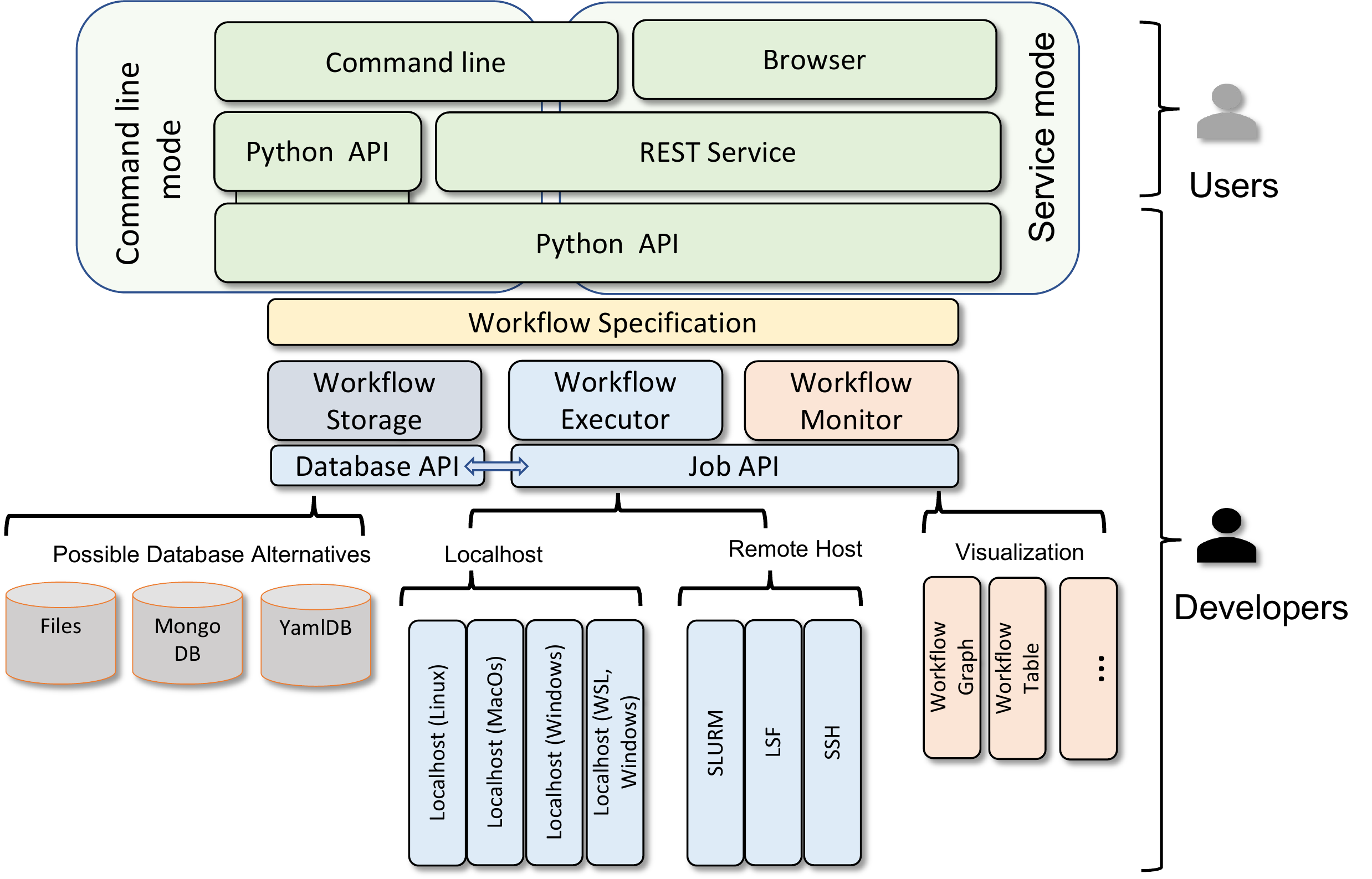}
}
\caption{Architecture of the Cloudmesh-cc Framework}\label{fig:arch}
\end{figure}

To fulfill our requirements, we have developed a framework for
workflow-controlled computing called Cloudmesh compute cluster,
otherwise referred to as {\em Cloudmesh-cc}. The architecture of the
framework is depicted in Figure~\ref{fig:arch}.

The framework is based on a layered architecture so it can be improved
and expanded on at each layer. We distinguish two kinds of users,
developers and end users, that define their workflows to run on
compute resources. For the end user, we provide a command line and a
very simple browser interface. Simplicity is important as it does not
require spending exorbitant amounts of time to learn how to use the
workflow framework. To support developers, we have designed a minimal
Python API that is also used to implement a REST Service.  Currently,
the REST Service is supposed to run on the client, but a service
deployment could also be conducted while assuring that proper
authentication and authorization are used. This is easily possible as
we use a well-known REST service framework (FastAPI) that can
integrate with common security frameworks to secure services. The API,
workflow specification, command line interface, and browser are well
documented with the help of Sphinx and OpenAPI.

The workflow specification plays an important role in not only
defining a workflow but also keeping the status of a currently
executed workflow. Here we have completely separated the status of the
workflow and synchronize the state of the workflow with pull requests
to the service that executes the computation. This allows the system
to be shut down at any time while the running jobs are completely
independent of the client application accessing the state. Thus, the
client appears to be stateless and fetches the state of the submitted
jobs on demand. It will return the latest state found on the job
execution services.

The workflows are defined with YAML. The workflow is stored in a
database that can be implemented using a variety of backends such as
YamlDB (which is file-based), MongoDB, or pickle. In our current
implementation, we simply use a file-based implementation as it is not
required to set up and manage more complex databases. This is an
important capability as we found that scientists and beginning
students do not want to engage in the hassle of setting up and
managing a datastore such as MongoDB.

As we use a YAML file to represent the status of the workflow, it is
easy to create monitoring components (for example, as part of a Web
browser). Various sophisticated graph display frameworks could be
used. For now, we have simply exposed the graph in table format using
datatables.net and the graph as SVG while leveraging Graphviz.

One of the most important parts of the framework is how we manage jobs
and monitor their status. For this, we have introduced an abstract job
class that is integrated into our workflow class. The job class can
define jobs, start them, and cancel them, to name only the most
important management methods. However, each job defines a status file
in which the actual progress of the job is recorded. This status file
is managed on the compute resource where the job is run and is queried
on demand to return the status to the client. This way, the status of
all jobs can be monitored. As our goal is not to run jobs that execute
in milliseconds, but rather in the second range, such status reporting
and propagation is well-suited for us. We have defined a special
status progress update specification that is universally
applicable. These jobs can be bash scripts, Python scripts, Jupyter
notebooks, or Slurm scripts.

Workflows are compilations of jobs to be run on nodes. These workflows
report information on the status of jobs as they are run, whether
locally or remotely. These types of jobs can be mixed with others into
a single workflow. This allows us, as already mentioned, to view the
progress of a workflow as it is being run in quasi-realtime.  The
workflow can be monitored as a graph or a DataTable. These interfaces
report the statuses of the jobs. Additionally, the order in which the
jobs are run can be specified, enabling prerequisite jobs and the
segmentation of a workflow.

To address the requirement of simplicity, the overall code is less
than 3600 lines of code, with additional 2600 lines for extensive test
cases.

The framework is managed as an open-source repository in GitHub and
uses Python as the implementation language. The code is compatible
with Windows, macOS, and Linux.

\FILE{specifying-workflows.tex}

\section{Specifying Workflows}\label{specifying-workflows}

The workflow definition for cloudmesh is rather simple and intuitive.
An example is provided in Figure~\ref{fig:workflow-example}. Here, a
DAG with three nodes are specified ($start \rightarrow fetch-data
\rightarrow compute \rightarrow analyze \rightarrow end$). The
workflow executes three scripts
(test-[fetch-data,compute,analyze].sh). It contains a specific start
and end node.

\subsection{Dependencies}

Each dependency is specified in the {\em dependency} section while
providing sequences of names in a list such as
{\scriptsize\texttt -start,fetch-data}
or, in our case where 3 more nodes are defined,
additional nodes are appended with commas, as in
{\scriptsize\texttt ,compute,analyze,end}.

In case one wants to execute nodes in parallel, we can simply define
them through a list such as

{\scriptsize
\begin{verbatim}
  dependencies:
    - start,a,end
    - start,b,end
    - start,c,end
\end{verbatim}
}

if the names of the nodes were $a$, $b$, and $c$. 

\subsection{Nodes}

Nodes can be customized in various ways within the workflow
configuration YAML file, including their job types (python, sh,
jupyter, or slurm), their virtual Python environment (by specifying
{\scriptsize\texttt venv}), their appearance on the graph, and other
characteristics. This is controlled through a number of attributes
used by the nodes in the DAG. The attributes are summarized in
Table~\ref{tab:node-attributes}.

\begin{figure}[htb]
{\scriptsize
\begin{verbatim}
workflow:
  nodes:
    start:
       name: start
    fetch-data:
       name: fetch-data
       user: gregor
       host: localhost
       kind: local
       status: ready
       label: '{name}\nprogress={progress}'
       script: test-fetch-data.sh
    compute:
       name: compute
       user: gregor
       host: localhost
       kind: local
       status: ready
       label: '{name}\nprogress={progress}'
       script: test-compute.sh
    analyze:
      name: analyze
      user: gregor
      host: localhost
      kind: local
      status: ready
      label: '{name}\nprogress={progress}'
      script: test-analyze.sh
    end:
       name: end
  dependencies:
    - start,fetch-data,compute,analyze,end
\end{verbatim}}
\caption{Workflow YAML Configuration file.}\label{fig:workflow-example}
\label{fig:yaml-file}
\end{figure}

\begin{table}[htb]
\caption{Node attributes.}\label{tab:node-attributes}
\resizebox{1.0\columnwidth}{!}{
\begin{tabular}{|l|p{6cm}|}
\hline
{\bf Attribute} & {\bf Description} \\
\hline
\hline
name &  A unique name of the job, must be the same as defined in the : line \\
\hline
user &  The username for the host \\
\hline
host &  The hostname \\
\hline
kind &  The kind of the job, which can be local, ssh, wsl, or slurm  \\
\hline
status &  The status of the job in integer value between 0 and 100 \\
\hline
label &  A custom-designed label \\
\hline
script &  The script name to be executed \\
\hline
\end{tabular}
}
\end{table}

\subsection{Node labels}

One particular useful attribute is that of a label. If no label is
used, the name of the node is used as the label. However, if a label
is specified, one can also use attribute names and timers to create
labels with implicit state information. This is done by introducing
variables through curly braces when defining the labels inside the
label defined in nodes within the YAML workflow file.

\begin{figure}[htb]
\smallskip
    {\scriptsize
    \begin{verbatim}
    workflow:
      nodes:
        start:
          label: 'start\n
                  Created={created.%Y/%m/%d, %H--%M--%S}\n
                  Workflow Started={t0.%Y/%m/%d, %H--%M--%S}\n
                  Now={now.%Y/%m/%d, %H--%M--%S}\n
                  Elapsed={dt0.%M--%S}'
    \end{verbatim}}
    \caption{Example of job label in YAML configuration file.}
    \label{fig:label-job}

\centering
\includegraphics[width=0.7\columnwidth]{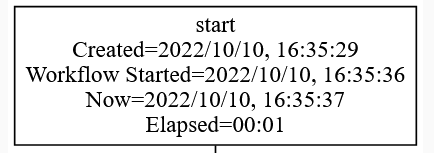}
\caption{An example node with labels.}\label{fig:label-node}
\end{figure}

For example, a label could be defined as showcased in
Figure~\ref{fig:label-job}. The appropriate values will be dynamically
replaced during the execution of the workflow.  This creates a node on
the graph that looks similar to the node showcased in
Figure~\ref{fig:label-node}.

Initially, the created and elapsed labels are {\scriptsize\texttt N/A}
if the workflow has not yet started, but they are replaced during
runtime. This can be observed by running a workflow in graph view in
the web interface.

In this format, we must use two dashes {\scriptsize\texttt --} to
separate the various components. However, when rendered, the dashes
will be replaced with a colon. Thus you can easily use years, months,
days, hours, minutes, and seconds can be arranged as desired, as long
as the corresponding letters remain consistent:
({\scriptsize\texttt \%Y} 
{\scriptsize\texttt \%m}
{\scriptsize\texttt \%d} 
{\scriptsize\texttt \%H} 
{\scriptsize\texttt \%M} 
{\scriptsize\texttt \%S}).
The time format must be specified immediately following the period
after a format-supported time variable.  If no format is specified
following the period after the variable, the datetime defaults to the
American format.  See Table~\ref{fig:labels-list} for a summary of
options for defining time based attributed to being replaced in the
label.

Nodes can be customized in various ways within the workflow
configuration YAML file, including their job types (python, sh,
jupyter, or slurm), their virtual Python environment (by specifying
{\scriptsize\texttt venv}), their appearance on the graph, and other
characteristics.

\begin{table}[htb]
\caption{List of possible labels for nodes on the graph.}
\label{fig:labels-list}

{\footnotesize
  \begin{tabular}{|p{3.5cm}|p{4cm}|}
    \hline
    {\bf Name} & {\bf Description} \\
    \hline
    \hline
    progress &  progress of job from 0-100 \\
    \hline
    now & current time \\
    \hline
    now.\verb|%Y%m%d,%H--%M--%S| &
   now in particular format (this can be used for other times as well) \\
    \hline
    created & time when workflow was created \\
    \hline
    t0.\verb|%Y%m%d,%H--%M--%S| &  workflow start time \\
    \hline
    t1.\verb|%Y%m%d,%H--%M--%S| & workflow end time \\
    \hline
    dt0.\verb|%Y%m%d,%H--%M--%S| & elapsed time since workflow began \\
    \hline
    dt1.\verb|%Y%m%d,%H--%M--%S| & total time of workflow once complete \\
    \hline
    tstart.\verb|%Y%m%d,%H--%M--%S| & job start time \\
    \hline
    tend.\verb|%Y%m%d,%H--%M--%S| & job end time \\
    \hline
    modified.\verb|%Y%m%d,%H--%M--%S| & job modified time \\
    \hline
    os. & operating system environment variable (like os.HOME) \\
    \hline
    cm. & cloudmesh variable that is read from \verb|cms set| \\
    \hline
\end{tabular}
}
\end{table}

\subsection{Shapes and Styles}

As we use graphviz for rendering, we have also added the ability to
change the shape and style of each node in the graph. The
available \href{https://graphviz.org/doc/info/shapes.html}{shapes}
and \href{https://graphviz.org/docs/attr-types/style/}{styles} are
listed in the Graphviz documentation \cite{www-graphviz}.

Figure~\ref{fig:shape-style-yaml} is an example of a node in YAML
format that uses a box shape and an empty style. The empty style
defaults to \verb|filled|, which allows the node to change
color when the job status is changed.

\begin{figure}[htb]
{\scriptsize
\begin{verbatim}
    workflow:
      nodes:
        start:
          label: 'start\n
                  Created={created.%Y/%m/%d, %H--%M--%S}\n
                  Workflow Started={t0.}\n
                  Elapsed={dt0.}'
          kind: local
          user: grey
          host: local
          status: ready
          exec: 'echo hello'
          name: start
          shape: box
          style: ''
\end{verbatim}}
\caption{Example of YAML config file that uses shape and style.}
\label{fig:shape-style-yaml}
\end{figure}

\subsection{Reporting Progress}\label{reporting-progress}

When running scripts/jobs inside a workflow, the scripts must leverage
some format of {\em cloudmesh.progress} to notify the user and the
backend client monitoring system. If progress is not reported, the
Workflow class cannot tell if the scripts are done.

The examples that are provided with cloudmesh-cc are already augmented
with {\em cloudmesh.progress}. Thus, if a user is running jobs through
cloudmesh cc workflows, they must integrate progress strings into the
log files that are monitored. This is available for shell, batch,
Python, and Jupyter scripts.

For shell and Slurm scripts, the script must contain a progress update
lines as follows:

{\scriptsize
\begin{verbatim}
  echo "# cloudmesh status=running progress=1 pid=$$"
\end{verbatim}
}

at the beginning of the script, and

{\scriptsize
\begin{verbatim}
  echo "# cloudmesh status=done progress=100 pid=$$"
\end{verbatim}
}

at the end of the script.

For Python scripts and Jupyter notebooks, it is easiest to use our
built-in progress method and import it from the cloudmesh.common
module as the example in Figure~\ref{fig:py-progress}.

\begin{figure}[htb]

{\scriptsize
\begin{verbatim}
  from cloudmesh.common.StopWatch import progress
  from cloudmesh.common.Shell import Shell
  filename = Shell.map_filename('./py_script.log').path
  progress(progress=1, filename=filename)
  ... execute your analysis here ...
  progress(progress=100, filename=filename)
\end{verbatim}
}

\caption{Progress report with the cloudmesh Python API.}
\label{fig:py-progress}

\end{figure}

When recording progress, the progress should be ascending between 1
and 100. The last value must report 100 or the node will not be
completed and the workflow gets stuck. The output of the progress will
be written into a {\scriptsize \verb|{workflowname}.log|} file that
will be probed continuously by the client to report the progress to
the user.

\subsection{Generating Progress Image as Graph}

The framework has a built-in capability to export the progress of the
workflow in a DAG.

\begin{figure}[htb]
\resizebox{1.0\columnwidth}{!}{
\begin{minipage}[b]{0.18\textwidth}
Definition and start \\
\includegraphics[height=9cm]{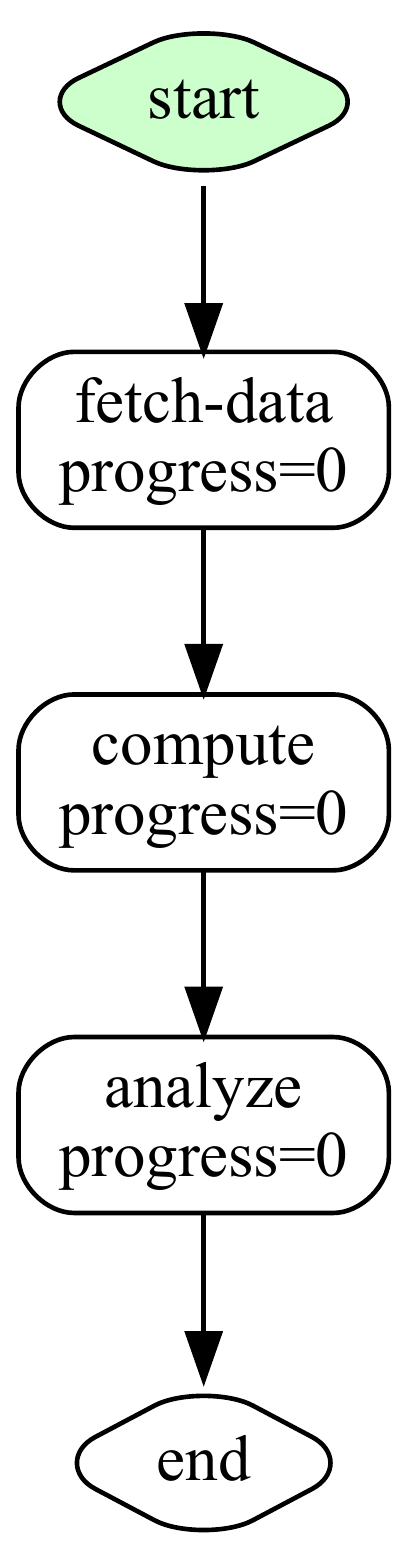}
\end{minipage} \ \
\begin{minipage}[b]{0.18\textwidth}
Running 1st task \\
\includegraphics[height=9cm]{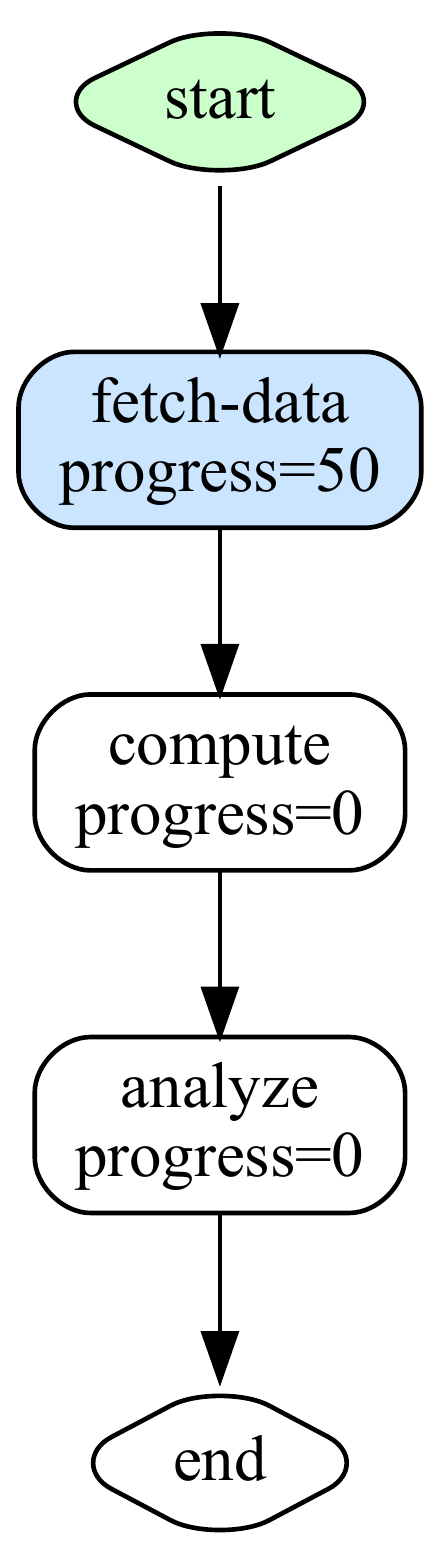} 
\end{minipage} \ \
\begin{minipage}[b]{0.18\textwidth}
Complete 1st task \\
\includegraphics[height=9cm]{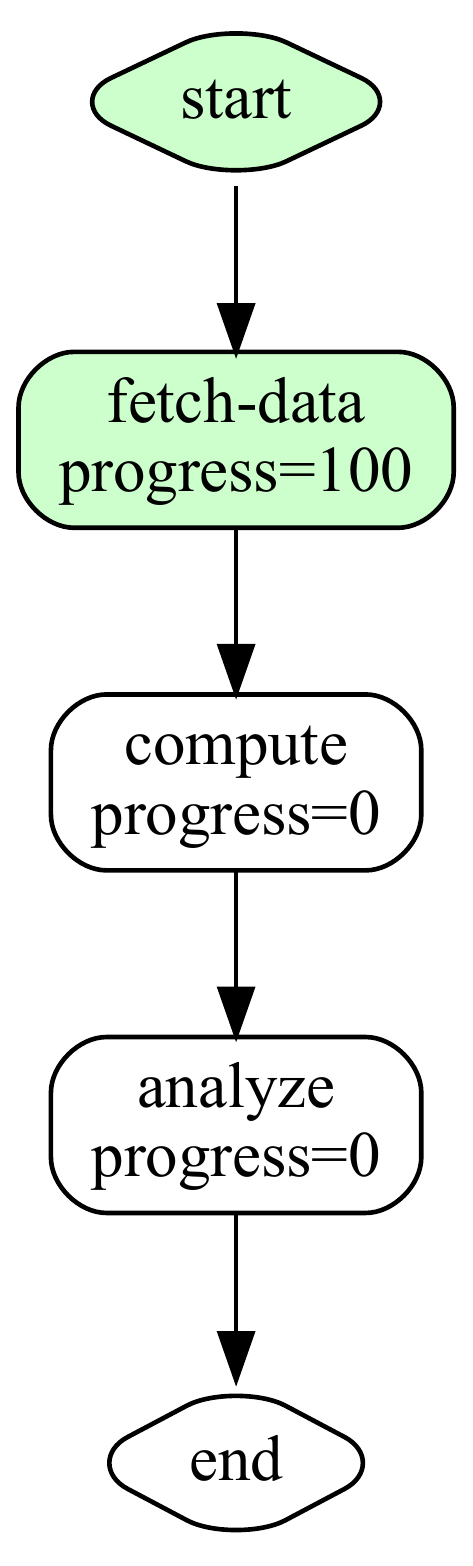}
\end{minipage} \ \
\begin{minipage}[b]{0.18\textwidth}
Complete 2nd task \\
\includegraphics[height=9cm]{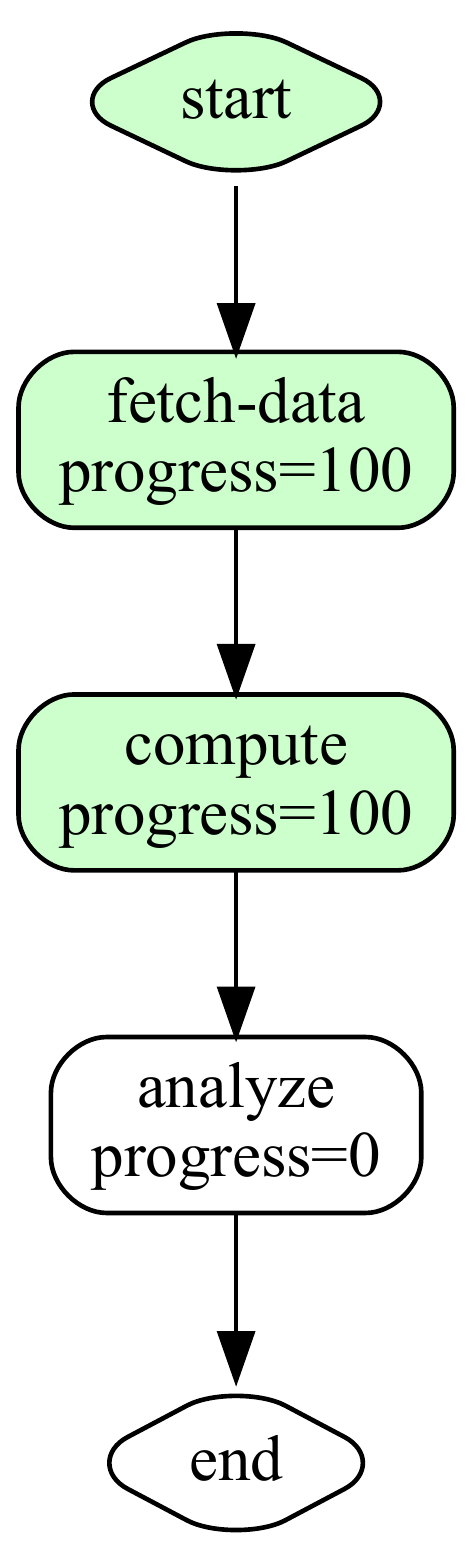}
\end{minipage} \ \
\begin{minipage}[b]{0.18\textwidth}
Complete workflow \\
\includegraphics[height=9cm]{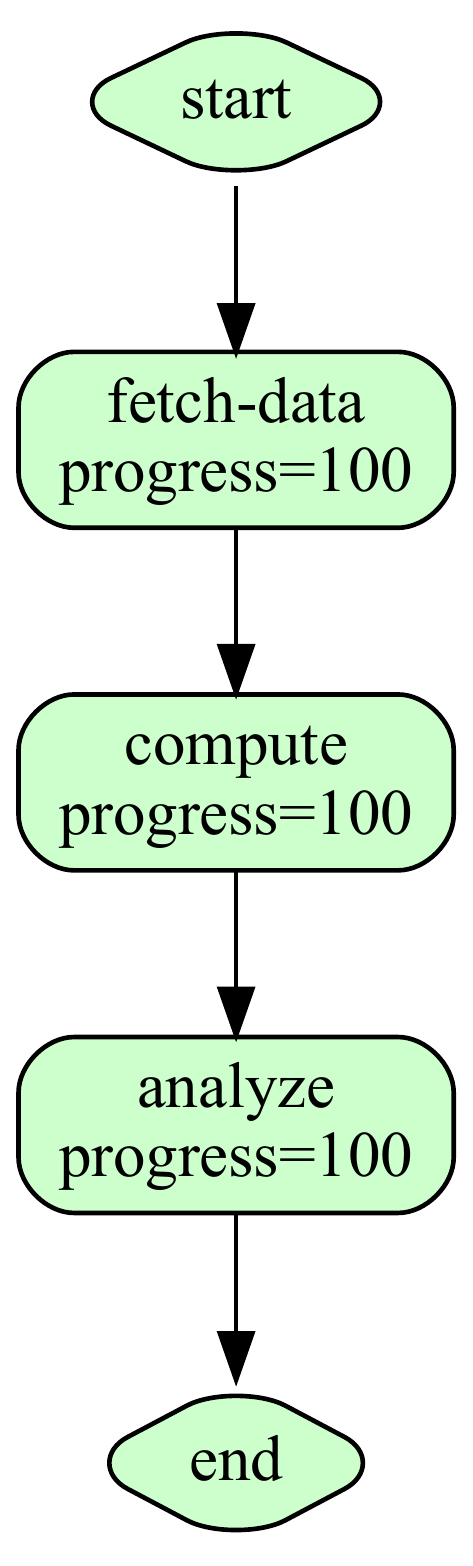}
\end{minipage} 
}
\caption{The gradual process of a simple workflow.}\label{fig:workflow-process}
\end{figure}

In Figure~\ref{fig:workflow-process}, we depict a simple example
workflow to be executed, where each task is executed sequentially. The
status of the execution can be displayed as a table or as a graph. In
Figure~\ref{fig:workflow-process}, we showcase how the graph changes
its appearance over time while using no label in the start and end
node and a defined label in the nodes for fetch-data, compute, and
analyze:

{\scriptsize
\begin{verbatim}
  label: "{name}\nprogress={progress}"
\end{verbatim}
}

\FILE{interfaces.tex}

\section{Cloudmesh-cc Interfaces}

In this section, we explain the various ways of interfacing with
cloudmesh-cc that are part of our design (see Figure~\ref{fig:arch}).

It is important to note that we have a {\em command line mode} that
interfaces directly with the backends while not requiring a
service. This includes an easy-to-use Python API.

In addition, we have used this API to implement a {\em service mode}, so we can
stand up a REST service as well as a GUI that can be accessed through a Web browser.
Please note that the service mode can also be accessed through the command line
in a terminal. To distinguish how we operate cloudmesh-cc, we use the term
{\em mode} to delineate it from a command that is entered in a terminal to
query the status of the workflow.

We will now discuss these interfaces in more detail and also showcase
how we can access them.

\subsection{Python API}

Cloudmesh-cc is implemented in Python. Extensive documentation is
available in GitHub and, using GitHub Actions, it is automatically
updated~\cite{github-cloudmesh-cc}. We distinguish two main classes
that are easy to use. The first is a {\em Job class} that can be
adapted to include new computational resource types for executing
jobs. The second is a {\em Workflow class} to coordinate the execution
of multiple jobs. Selected methods of the Job class include are listed
in Figure~\ref{fig:code-job}. Important to note is that the code for
the scripts are all managed locally, and a synchronization step is
invoked prior to one running the job to assure that the latest script
and code to be executed on the compute resource is available.

The most important part of using the Workflow Python API is showcased
in Figure~\ref{fig:code-job}, where we explain how easy it is to set
up a workflow with the Python API.

\begin{figure}[!h]
{\scriptsize
\begin{verbatim}
class Job:
   ...
   def clear():
      """Clear job progress."""
   def create(filename=None, script=None, exec=None):
      """Create a template for the script with progress."""
   def get_log(refresh=True):
      """Get the log of the job."""
   def get_pid(refresh=False):
      """Get the pid that the job is running within determined by 
         the compute resource it is running on."""
   def get_progress(refresh=False):
       """Get the progress of the job from the compute resource."""
   def get_status(refresh=False):
       """Get the status of the job."""
   def kill():
       """Kill the job."""
   def run():
       """Run the job."""
   def sync():
       """Synchronise the current directory with the remote. Copies 
          the shell script to the experiment directory and ensures 
          that the file is copied with the sync command."""
   def watch(period=10):
       """Watch the job and check for changes in the given period."""
\end{verbatim}}
\caption{Pseudo code for the Job class with selected methods.}
\label{fig:code-job}

\bigskip

{\scriptsize
\begin{verbatim}
class Workflow:
    ...
    def add_dependencies(dependency):
        """Add a job dependency to the workflow (and the graph)."""
    def add_dependency(source, destination):
        """Add a job dependency to the workflow (and the graph)."""
    def add_job( ... ):
        """Add a job to the workflow with appropriate parameters."""
    def display(filename=None, name='workflow', first=True):
        """Show the graph of the workflow."""
    def job(name):
        """Return the details of a job within the workflow."""
    def load(filename, clear=True):
        """Load the workflow."""
    def remove_job(name, state=False):
        """Remove a particular job from the workflow."""
    def remove_workflow():
        """Delete workflow from the local file system."""
    def run_parallel( ... ):
        """Run a workflow in a parallel fashion."""
    def run_topo(order=None, dryrun=False, show=True, filename=None):
        """Run the workflow in a topological order."""
    def save(filename=None):
        """Save the workflow."""
    def save_with_state(filename, stdout=False):
        """Save the workflow with state."""
    def sequential_order():
        """Return a list of the topological order of the workflow."""
    property table:
        """Return a table of the workflow."""
    def update_progress(name):
        """Manually update the progress of a job according to its log 
           file."""
    def update_status(name, status):
        """Manually update a job’s status."""
\end{verbatim}}
\caption{Pseudo code for the Job class with selected methods.}
\label{fig:code-workflow}
\end{figure}

\begin{figure}[htb]
{\scriptsize
\begin{verbatim}
    w = Workflow(name="workflow-analyze")
    # load a preexisting workflow
    # w.load(filename="source.yaml")
    
    # add jobs and dependencies explicitly
    w.add_job(name="fetch-data",
              exec="hostname",
              host="supercomputer-a", # defined in .ssh
              label="{name}\nprogress={progress}",
              kind="local",
              status="ready",
              progress=0)
    w.add_job(name="compute", command="... TBD ...")
    w.add_job(name="analyze", command="... TBD ...")
    w.add_dependencies(
              dependency="start,fetch-data,compute,analyze")
    w.run_topo()
\end{verbatim}}
\caption{Pseudo code for the Job class with selected methods.}
\label{fig:code-workflow-example}

\bigskip

{\scriptsize
\begin{verbatim}
      cms cc workflow add [--name=NAME] [--job=JOB] ARGS...
      cms cc workflow add [--name=NAME] --filename=FILENAME
      cms cc workflow delete [--name=NAME] [--job=JOB]
      cms cc workflow list [--name=NAME] [--job=JOB]
      cms cc workflow run [--name=NAME] 
                          [--job=JOB] 
                          [--filename=FILENAME]
      cms cc workflow [--name=NAME] --dependencies=DEPENDENCIES
      cms cc workflow status --name=NAME [--output=OUTPUT]
      cms cc workflow graph --name=NAME
\end{verbatim}}

\caption{Command line interface to the workflow in terminal mode.}
\label{fig:code-workflow-commandline}

\bigskip

{\scriptsize
\begin{verbatim}
      cms cc start [-c] [--reload] [--host=HOST] [--port=PORT]
      cms cc stop
      cms cc status
      cms cc workflow service add [--name=NAME] FILENAME
      cms cc workflow service list [--name=NAME] [--job=JOB]
      cms cc workflow service add [--name=NAME] [--job=JOB] ARGS...
      cms cc workflow service run --name=NAME
\end{verbatim}}

\caption{Command line interface to the workflow in service mode.}
\label{fig:code-workflow-service-commandline}.


{\centering
\includegraphics[width=0.52\columnwidth]{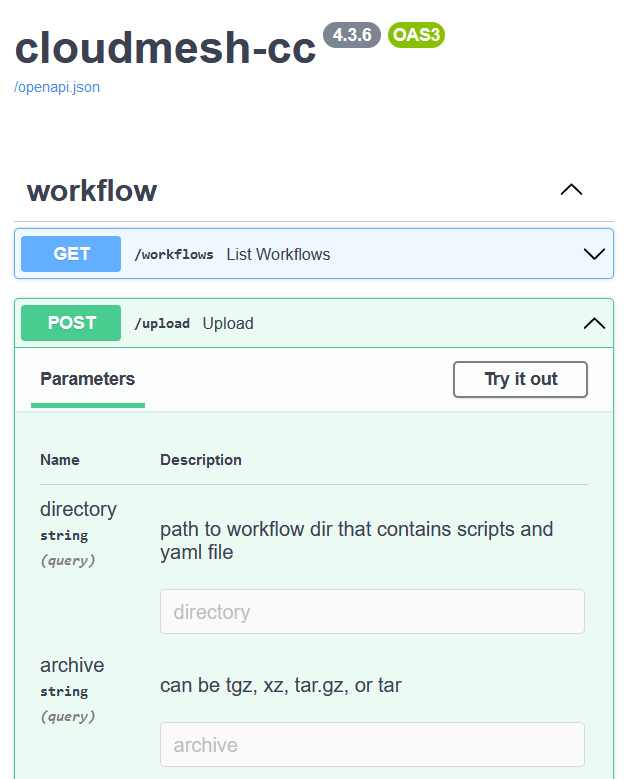}
}
\caption{Browser API GUI for Cloudmesh Compute Cluster.}

\label{fig:openapi}

\end{figure}

\subsection{Command Line Mode}

The command line mode is an implementation that does not use a backend
service. It runs in the terminal until the workflow is completed. All
states of the jobs are managed on the compute service on which the job
is run and the status is replicated into the client on demand. This
allows easy reporting of the status through a table and a graph
display using client-based rendering tools. The various commands to
interact with workflows on the command line are shown in
Figure~\ref{fig:code-workflow-commandline}.

\subsection{Service Mode}

The Python API has been used to implement a REST service. To easily
interact with the REST service we have also added a couple of
convenient commands one can issue from a terminal. They are listed in
Figure~\ref{fig:code-workflow-service-commandline}. It is clear from
these methods that starting, stopping, and getting the status of the
service is very easy. In contrast to the command line mode, this
service mode has an additional keyword {\em service} to interact with
the rest service.

Certainly one can also directly interact with the REST service API for
this service. For ease of use, we have exposed the interface through
an OpenAPI specification that is available to the user as shown in
Figure~\ref{fig:openapi}.

Thus, other tools such as curl or other languages supporting URL
requests can be used. A curl example to list the workflow
specifications uploaded to the service is as follows:

{\scriptsize
\begin{verbatim}
$ curl -X 'GET' 'http://127.0.0.1:8000/workflows' -H 'accept: application/json'
\end{verbatim}}

As we use a REST service, we can also easily upload the workflow
through a Python-enabled REST call. We will use Python requests to
demonstrate this upload feature. To showcase the various ways to
access the service, we focus on uploading a tar file that contains the
workflow as well as the scripts or programs used in the workflow. The
tar file is called {\em workflow-example.tar}.

To also allow programming against the REST service in Python, a Python
API similar to that of the command line mode is
available. Figure~\ref{fig:code-workflow-rest-commandline} showcases
this API.

To simplify interaction with the REST service, we have created a special
{\em RESTWorkflow class} that is similar to the command module API,
but instead uses the REST interface to the service rather than direct
communication with the command client API.

The user can also simply use the requests module in Python to
interface with the API. Figure~\ref{fig:code-workflow-requests}
demonstrates how to use requests to upload a workflow by using an
archive file that contains the YAML configuration file and the
scripts.

\begin{figure}[t]

{\scriptsize
\begin{verbatim}
class RESTWorkflow
    ...
    def add_job(workflow_name, **kwargs):
        """Add a job to the workflow."""
    def delete_workflow(workflow_name, job_name=None):
        """Delete a workflow by using REST."""
    def get_workflow(workflow_name, job_name=None):
        """Retrieve a workflow by using REST."""
    def list_workflows():
        """Return a list of workflows that is found within
           the server."""
    def run_workflow( ... ):
        """Run a workflow by using REST."""
    def upload_workflow( ... ):
        """Upload a workflow by using REST."""
\end{verbatim}}

\caption{Pseudo code for the Job class with selected methods.}
\label{fig:code-workflow-rest-commandline}

\bigskip

{\scriptsize
\begin{verbatim}
  import requests
  r = requests.post(
      'http://127.0.0.1:8000/workflow?archive=workflow-example.tar')
  print(r.text)
\end{verbatim}}

\caption{Upload to the REST service with Python requests.}
\label{fig:code-workflow-requests}

\bigskip

{\scriptsize
\begin{verbatim}
  $ curl -X 'POST' 'http://127.0.0.1:8000/ 
                    workflow?archive=workflow-example.tar'
         -H 'accept: application/json' -d ''
\end{verbatim}}

\caption{Upload to the REST service with curl.}
\label{fig:code-workflow-curl}

\end{figure}

\subsection{Webservice GUI}

A convenient Web service is included in Cloudmesh cc. It allows the
user to manage and visualize the status of workflows through a Web
browser interface. At this time, the focus is that the interface can
be run by a single user on the local machine. This allows remote
executions of workflow nodes run completely independent from cloudmesh
cc and interaction is possible in asynchronous mode.

As the service is using also an OpenAPI 2.0 specification, the
workflow can also be uploaded implicitly through the specification
GUI. Navigate to {\scriptsize \texttt{http://127.0.0.1:8000/docs}} and
use the POST Upload method. Then click {\scriptsize \texttt{Try\ it\ out}}
and enter the location of the tar file, followed by clicking
{\scriptsize \texttt{Execute}}.

Obviously, any REST service or REST API can be used, allowing the user
to interface to it from different programming languages or frameworks.

The web server provides a more customizable, easy-to-use interface for
the Workflow class which can be started, viewed, and stopped with the
appropriate command suc as

{\scriptsize
\begin{verbatim}
  $ cms cc start
  $ cms cc view
  $ cms cc stop
\end{verbatim}}

The view can also be achieved by opening the 
link \newline {\scriptsize \texttt{http://127.0.0.1:8000/}}.

The browser provides an interface to view preexisting workflows in
both a DataTable format and as a graph format. Both views will update
in a live, automatic fashion as the workflows are run, reporting
dynamic job status and progress.

\begin{figure}[htb]
{\centering
  \includegraphics[width=1.0\columnwidth]{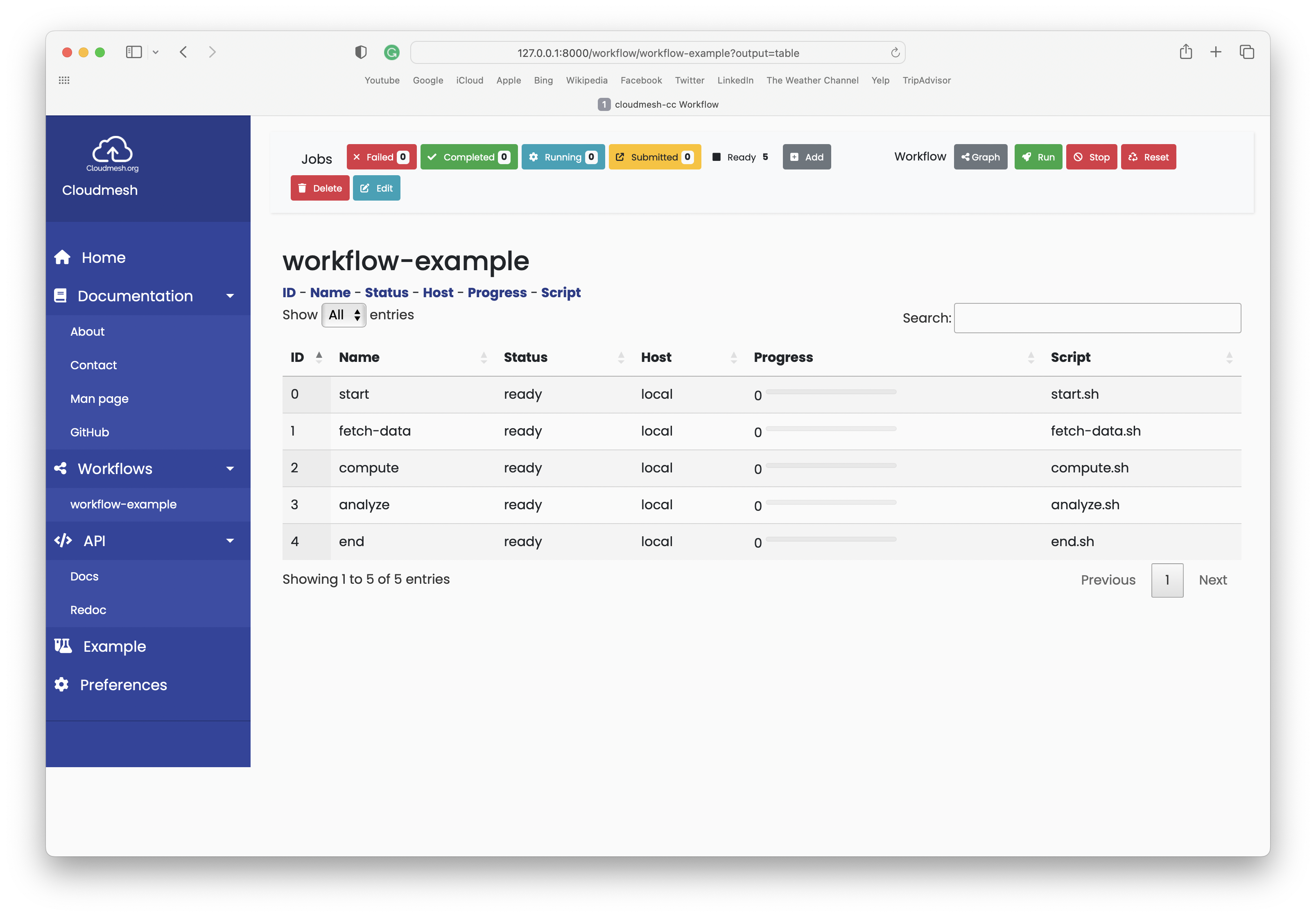}
}
\vspace{-1.0cm}
\caption{Cloudmesh cc workflow table view.}\label{fig:table}

\bigskip
{\centering
  \includegraphics[width=1.0\columnwidth]{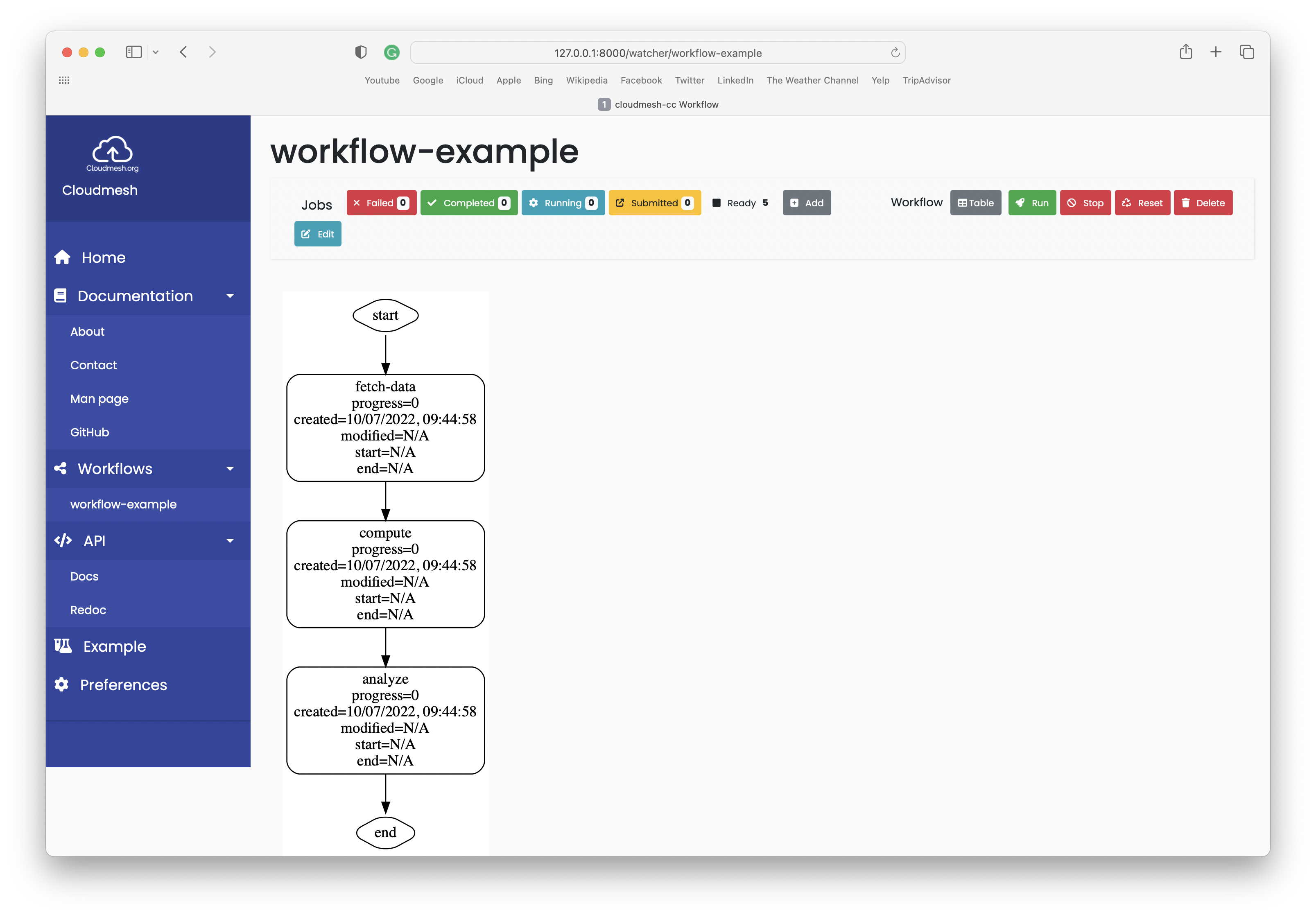}
 }
\vspace{-1.0cm}
 \caption{Cloudmesh cc workflow graph view.}

\label{fig:graph}
\end{figure}

For a quick and easy example of leveraging this GUI interface, click
on the Example tab in the left-hand sidebar. Then, a workflow-example
will appear underneath Workflows. Click on the workflow-example and
run the workflow by clicking the green Run button in the top-right. As
the workflow runs, the user is able to click on the Graph button to
view the graph interface (see Figure~\ref{fig:graph}) and back to the
Table button for the table interface (see Figure~\ref{fig:table}), as
desired, to view the workflow's progression.

\subsection{Run the workflow}\label{run-the-workflow}

The workflow can be run easily via the GUI. We added a special set of
buttons to the workflow table and graph display to simplify running of
the workflow. Certainly, the workflow can also be activated while
calling the appropriate REST call, either through Python, the OpenAPI
docs page, or, for example, a curl call.
Figures~\ref{fig:workflow-run-a}-\ref{fig:workflow-run-c}
showcase various methods to run the example workflow, which is called
{\em workflow-example}.

\begin{figure}[htb]

{\scriptsize\begin{verbatim}
  $ curl -X 'GET' 'http://127.0.0.1:8000/
                   workflow/run/workflow-example?show=True'
         -H 'accept: application/json'
\end{verbatim}}

\caption{Running the example workflow with curl.}
\label{fig:workflow-run-a}
\bigskip

{\scriptsize\begin{verbatim}
  from cloudmesh.cc.workflowrest import RESTWorkflow
  rest = RESTWorkflow()
  result = rest.run_workflow('workflow-example')
\end{verbatim}}

\caption{Running the example workflow with cloudmesh RESTWorkflow API.}
\label{fig:workflow-run-b}
\bigskip

{\scriptsize\begin{verbatim}
  import requests
  url = f'http://127.0.0.1:8000/workflow/run/workflow-example?show=True'
  r = requests.get(url)
  print(r)
\end{verbatim}}

\caption{Running the example workflow with requests API.}
\label{fig:workflow-run-c}

\end{figure}

\FILE{cloudmesh.tex}

\section{Other Cloudmesh Features}

Cloudmesh comes with a sophisticated package management system,
allowing the integration of packages on demand targeting various providers
and capabilities including a built-in command shell (not just only
a command line tool). Cloudmesh was first developed as a hybrid cloud
API, command line and command shell framework. It provided interfaces
to AWS, Azure, Google, and OpenStack clouds for virtual
machine\footnote{cloudmesh also provided support for clouds that are no
longer supported such as Eucalyptus and Open Cirrus. Academic clouds
such as Chameleon Cloud were also supported.} and data file services. It
is characterized by defining default templates for virtual machine
management on these clouds. Hence, it was possible to switch between clouds
with only a few commands and stage virtual machines on them, such
as with the commands demonstrated in Figure~\ref{fig:cms}.

\begin{figure}[htb]

{\scriptsize\begin{verbatim}
  $ cms vm start --cloud aws
  $ cms vm start --cloud azure
\end{verbatim}}

\caption{Simple VM management for hybrid clouds}
\label{fig:cms}.
\end{figure}  

In addition, we have developed a package called GAS that addresses
the creation of analytics REST services from Python functions. This
package was developed to address the problem that integrating
deployment frameworks in the age of cloud computing is often out of
reach for domain experts. GAS is a simple framework allowing even
non-experts to deploy and host services in the cloud. To avoid vendor
lock-in, it supports multiple vendors through the use of cloudmesh vm
management~\cite{las21-gas}.

\FILE{applications.tex}

\section{Workflow Applications}

We have applied the workflow system on a number of applications. All
workflows for these applications are available in our GitHub and can
be adapted easily. This includes the Cloudmask and MNIST application
workflows which we describe next in more details.

\FILE{Cloudmask.tex}

\subsection{MLCommons Cloudmask Workflow}
\label{cloudmask-workflow}

\begin{figure*}[htb]
\centering
\includegraphics[width=0.75\textwidth]{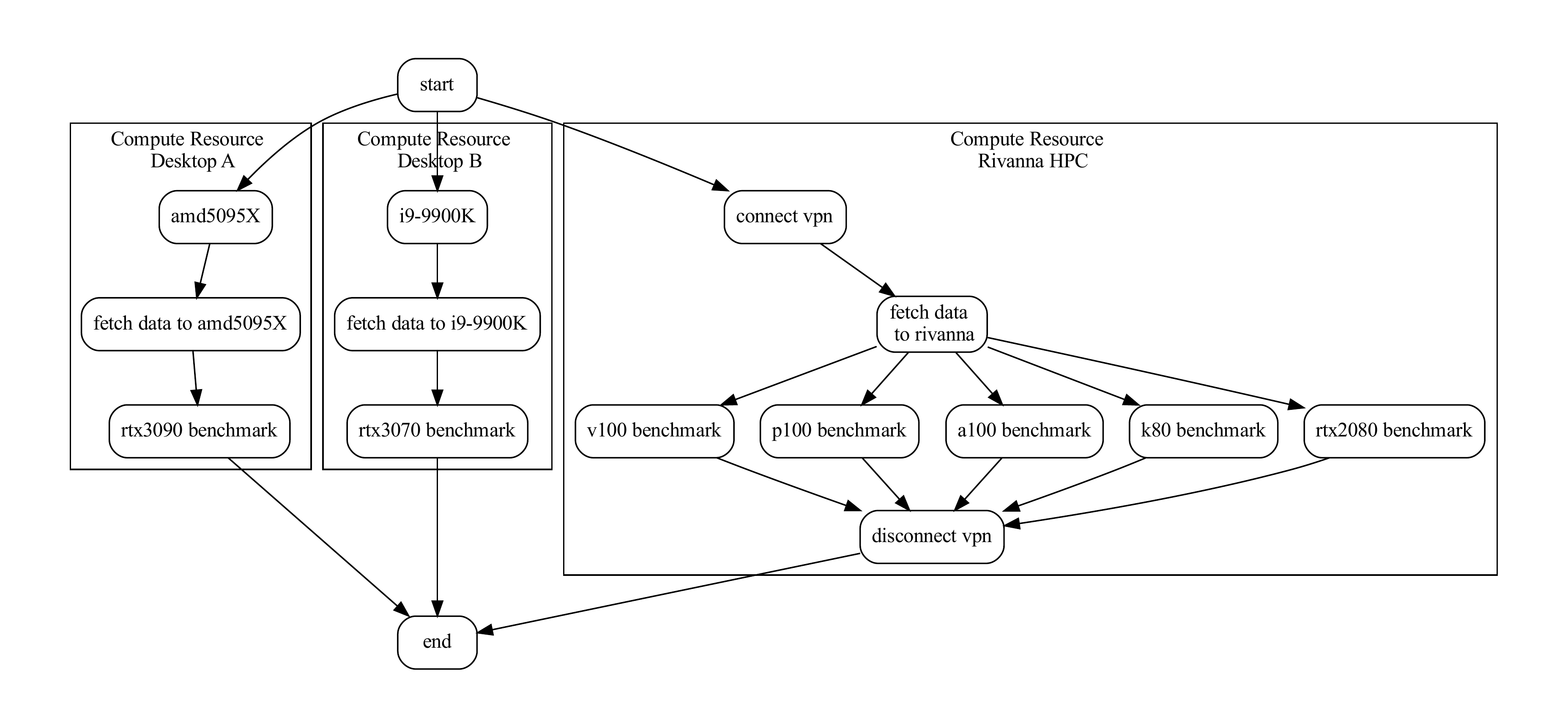}
\caption{Workflow for Cloudmask}\label{fig:cloudmaskwf}
\end{figure*}

Cloudmask is a program that develops a model to classify sections of
satellite images as either containing clouds or clear sky by using
machine learning. This is beneficial for temperature measurement and
meteorology. Information regarding Cloudmask can be found on its
GitHub page~\cite{www-cloudmask}. One of our goals is to run
Cloudmask for benchmarking. As benchmarking Cloudmask requires
several phases and scripts, including a mixture of shell scripts and
Python scripts, leveraging Cloudmesh-cc provides a much easier runtime
instead of manually issuing many commands at a terminal. We have
created a sample workflow the runs a coordinated workflow across a
number of hybrid resources. This includes an HPC computer at the
University of Virginia called Rivanna, as well as two desktop
computers. This workflow can easily be adapted to include other
machines. In this particular workflow, we execute the benchmarks on a
number of different CUDA cards (see Figure~\ref{fig:cloudmaskwf}. For
Rivanna, the code also utilizes our cloudmesh-vpn component that
provides the ability to connect to the UVA VPN from Python, then
fetches the data and executes the various benchmarks once the data is
available.

The workflow will take approximately 24 hours to run if resources are
available. The workflow iterates through the five GPUs available on
Rivanna, including V100, P100, A100, K80, and RTX2080, and runs the
program three times on each GPU. Each run trains the model with 10,
30, and 50 epochs for benchmarking. Upon completing a run, the logs
and benchmarks are written into a results folder. In the appendix, we
showcase how to run a portion of this workflow while utilizing
only Rivanna (see Appendix~\ref{sec:running-cloudmask}).

\FILE{mnist.tex}

\subsection{MNIST Workflow}\label{mnist-workflow}

MNIST is a well-known program to detect handwritten digits. It
provides value for our work because it is well understood and is used
in many educational efforts. Also, we created a workflow that
integrates the UVA Rivanna HPC (the workflow can easily be adapted to other
machines). The nice feature about this application is that it can be
configured to run very quickly while still using various GPUs and
benchmark their runtimes for running several MNIST Python
programs. These programs include machine learning processing,
convolutional neural network, long short-term memory, recurrent neural
network, and others. The programs can be found on
GitHub~\cite{www-mnist-programs}.

As for the workflow, we adapted it not only to run one algorithm but
multiple in an iteration across the GPUs (similar to
\ref{fig:cloudmaskwf}).

On a successful run, the output will be receiving runtimes similar to:

\begin{table}[!ht]
\caption{MNIST Performance as obtained by cloudmesh-cc on various graphics cards using workflow scheduling}
    \centering
    \begin{tabular}{lr}
    \hline
        Name & Time \\ \hline
        a100 & 106.046 \\ 
        v100 & 138.087 \\ 
        rtx2080 & 138.048 \\
        k80 & 171.057 \\ 
        p100 & 202.055 \\
    \end{tabular}
    \label{table:mnist-times}
  \end{table}

\FILE{conclusion.tex}

\section{Conclusion}

We have designed and implemented a Hybrid Reusable Computational
Analytics Workflow Management with the help of the cloudmesh component
framework. The component added focuses on the management of workflows
for computational analytics tasks and jobs. The tasks can be executed
on remote resources via ssh and even access queuing systems such as
Slurm. In addition, we can integrate the current computer on which the
workflow is running. This can include operating systems such as Linux,
macOS, Windows, and even Windows Subsystem for Linux. Through
cloudmesh, access to a command line and a command shell is provided. A
simple API and a REST interface are provided. The framework also has
an elementary Web browser interface that allows visualizing the
execution of the workflow. It is important to know that the workflow
can be started on remote resources and is running completely
independently from the client tool once a task is started. This allows
a ``stateless'' model that can synchronize with the remotely started
jobs on demand. Hence, the framework is self-recovering in case of
network interruptions or power failure. Due to our experiences with
real (and many) infrastructure failures at the authors' locations, the
availability of such a workflow-guided system was
beneficial. Furthermore, the developed code is rather small and, in
contrast to other systems, is less complex. Hence, it is suitable for
educational aspects as it is used for master's and undergraduate level
research projects. The project has also been practically utilized
while generating benchmarks for the MLCommons Science Working Group
showcasing real-world applicability beyond a student research project.

\FILE{acknowledgements.tex}


\section*{Acknowledgements}

Continued work was in part funded by the NSF CyberTraining: CIC:
CyberTraining for Students and Technologies from Generation Z with the
award numbers 1829704 and 2200409 and NIST 60NANB21D151T. We would like to
thank the students that participated in the REU for their help and
evaluation in a pre-alpha version of the code. We are excited that
this effort contributed significantly to their increased understanding
of Python and how to develop in a team using the Python ecosystem.


\bibliographystyle{ACM-Reference-Format}
\bibliography{vonLaszewski-cloudmesh-cc}


\begin{thebibliography}{32}


\ifx \showCODEN    \undefined \def \showCODEN     #1{\unskip}     \fi
\ifx \showDOI      \undefined \def \showDOI       #1{#1}\fi
\ifx \showISBNx    \undefined \def \showISBNx     #1{\unskip}     \fi
\ifx \showISBNxiii \undefined \def \showISBNxiii  #1{\unskip}     \fi
\ifx \showISSN     \undefined \def \showISSN      #1{\unskip}     \fi
\ifx \showLCCN     \undefined \def \showLCCN      #1{\unskip}     \fi
\ifx \shownote     \undefined \def \shownote      #1{#1}          \fi
\ifx \showarticletitle \undefined \def \showarticletitle #1{#1}   \fi
\ifx \showURL      \undefined \def \showURL       {\relax}        \fi
\providecommand\bibfield[2]{#2}
\providecommand\bibinfo[2]{#2}
\providecommand\natexlab[1]{#1}
\providecommand\showeprint[2][]{arXiv:#2}

\bibitem[airflow(2022)]%
        {www-airflow}
airflow \bibinfo{year}{2022}\natexlab{}.
\newblock \bibinfo{title}{{Apache Airflow}}.
\newblock
\newblock
\urldef\tempurl%
\url{https://airflow.apache.org}
\showURL{%
\tempurl}
\newblock
\shownote{[Online; accessed 14. Oct. 2022]}.


\bibitem[{Amazon Web Services}(2022a)]%
        {www-aws-batch-workflow}
\bibfield{author}{\bibinfo{person}{{Amazon Web Services}}.}
  \bibinfo{year}{2022}\natexlab{a}.
\newblock \bibinfo{title}{{Encoding workflow dependencies in AWS Batch
  {$\vert$} Amazon Web Services}}.
\newblock
\newblock
\urldef\tempurl%
\url{https://aws.amazon.com/blogs/hpc/encoding-workflow-dependencies-in-aws-batch}
\showURL{%
\tempurl}
\newblock
\shownote{[Online; accessed 14. Oct. 2022]}.


\bibitem[{Amazon Web Services}(2022b)]%
        {www-aws-stepfunctions}
\bibfield{author}{\bibinfo{person}{{Amazon Web Services}}.}
  \bibinfo{year}{2022}\natexlab{b}.
\newblock \bibinfo{title}{{Serverless Workflow Orchestration AWS Step
  Functions}}.
\newblock
\newblock
\urldef\tempurl%
\url{https://aws.amazon.com/step-functions}
\showURL{%
\tempurl}
\newblock
\shownote{[Online; accessed 14. Oct. 2022]}.


\bibitem[{Amazon Web Services}(2022c)]%
        {www-aws-workflow}
\bibfield{author}{\bibinfo{person}{{Amazon Web Services}}.}
  \bibinfo{year}{2022}\natexlab{c}.
\newblock \bibinfo{title}{{What is a Workflow? Cloud Computing Workflows
  Introduction - AWS}}.
\newblock
\newblock
\urldef\tempurl%
\url{https://aws.amazon.com/what-is/workflow}
\showURL{%
\tempurl}
\newblock
\shownote{[Online; accessed 14. Oct. 2022]}.


\bibitem[{Amazon Web Services}(2022d)]%
        {www-aws-swf}
\bibfield{author}{\bibinfo{person}{{Amazon Web Services}}.}
  \bibinfo{year}{2022}\natexlab{d}.
\newblock \bibinfo{title}{{What is Amazon Simple Workflow Service? - Amazon
  Simple Workflow Service}}.
\newblock
\newblock
\urldef\tempurl%
\url{https://docs.aws.amazon.com/amazonswf/latest/developerguide/swf-welcome.html}
\showURL{%
\tempurl}
\newblock
\shownote{[Online; accessed 14. Oct. 2022]}.


\bibitem[Amin et~al\mbox{.}(2004)]%
        {las-2004-gridant}
\bibfield{author}{\bibinfo{person}{Kaizar Amin}, \bibinfo{person}{Mihael
  Hategan}, \bibinfo{person}{Gregor von Laszewski}, \bibinfo{person}{estor~J.
  NZaluzec}, \bibinfo{person}{Shawn Hampton}, {and} \bibinfo{person}{Albert
  Rossi}.} \bibinfo{year}{2004}\natexlab{}.
\newblock \showarticletitle{GridAnt: A Client-Controllable Grid Workflow
  System}. In \bibinfo{booktitle}{\emph{37th Hawai'i International Conference
  on System Science}}, Vol.~\bibinfo{volume}{7}. \bibinfo{publisher}{IEEE
  Computer Society, Los Alamitos, CA, USA}, \bibinfo{address}{Big Island, HW}.
\newblock
\showISSN{1530-1605}
\urldef\tempurl%
\url{https://doi.org/10.1109/HICSS.2004.1265491}
\showDOI{\tempurl}
\newblock
\shownote{The original paper is: von Laszewski, Gregor and Kaizar Amin and
  Shawn Hampton and Sandeep Nijsure. Technical report and Argonne National
  Laboratory and 31 July 2002.
  https://laszewski.github.io/papers/vonLaszewski-gridant.pdf}.


\bibitem[{Argonne National Laboratory}(2022)]%
        {www-parsl}
\bibfield{author}{\bibinfo{person}{{Argonne National Laboratory}}.}
  \bibinfo{year}{2022}\natexlab{}.
\newblock \bibinfo{title}{{Parsl: Parallel Scripting in Python}}.
\newblock
\newblock
\urldef\tempurl%
\url{https://parsl-project.org}
\showURL{%
\tempurl}
\newblock
\shownote{[Online; accessed 14. Oct. 2022]}.


\bibitem[Cloud(2022)]%
        {www-gcloud}
\bibfield{author}{\bibinfo{person}{Google Cloud}.}
  \bibinfo{year}{2022}\natexlab{}.
\newblock \bibinfo{title}{{Quickstart: Create a workflow using the gcloud CLI
  {$\vert$} Workflows {$\vert$} Google Cloud}}.
\newblock
\newblock
\urldef\tempurl%
\url{https://cloud.google.com/workflows/docs/create-workflow-gcloud}
\showURL{%
\tempurl}
\newblock
\shownote{[Online; accessed 14. Oct. 2022]}.


\bibitem[Dario(2021)]%
        {www-azure-enterprise-workflow}
\bibfield{author}{\bibinfo{person}{F{\ifmmode\acute{a}\else\'{a}\fi}bio
  Dario}.} \bibinfo{year}{2021}\natexlab{}.
\newblock \bibinfo{booktitle}{\emph{{Enterprise Workflows with Azure Logic Apps
  and Azure Functions}}}.
\newblock \bibinfo{publisher}{gft-engineering}.
\newblock
\showISBNx{978-303971151}
\urldef\tempurl%
\url{https://medium.com/gft-engineering/enterprise-workflows-with-azure-logic-apps-and-azure-functions-303971a15f10}
\showURL{%
\tempurl}


\bibitem[ecfan(2022)]%
        {www-wsdl}
\bibfield{author}{\bibinfo{person}{ecfan}.} \bibinfo{year}{2022}\natexlab{}.
\newblock \bibinfo{title}{{Workflow Definition Language schema reference -
  Azure Logic Apps}}.
\newblock
\newblock
\urldef\tempurl%
\url{https://learn.microsoft.com/en-us/azure/logic-apps/logic-apps-workflow-definition-language}
\showURL{%
\tempurl}
\newblock
\shownote{[Online; accessed 14. Oct. 2022]}.


\bibitem[{GitHub}(2022)]%
        {www-github-rest-cancel}
\bibfield{author}{\bibinfo{person}{{GitHub}}.} \bibinfo{year}{2022}\natexlab{}.
\newblock \bibinfo{title}{{Cancel a workflow run - Workflow runs}}.
\newblock
\newblock
\urldef\tempurl%
\url{https://docs.github.com/en/rest/actions/workflow-runs#cancel-a-workflow-run}
\showURL{%
\tempurl}
\newblock
\shownote{[Online; accessed 14. Oct. 2022]}.


\bibitem[graphviz(2021)]%
        {www-graphviz}
graphviz \bibinfo{year}{2021}\natexlab{}.
\newblock \bibinfo{title}{{Graphviz}}.
\newblock
\newblock
\urldef\tempurl%
\url{https://graphviz.org}
\showURL{%
\tempurl}
\newblock
\shownote{[Online; accessed 21. Oct. 2022]}.


\bibitem[{IBM}(2021)]%
        {www-business-rest-ibm}
\bibfield{author}{\bibinfo{person}{{IBM}}.} \bibinfo{year}{2021}\natexlab{}.
\newblock \bibinfo{title}{{Business Automation Workflow REST APIs
  programming}}.
\newblock
\newblock
\urldef\tempurl%
\url{https://www.ibm.com/docs/en/baw/19.x?topic=apis-business-automation-workflow-rest}
\showURL{%
\tempurl}
\newblock
\shownote{[Online; accessed 14. Oct. 2022]}.


\bibitem[{IBM}(2022)]%
        {www-lsf}
\bibfield{author}{\bibinfo{person}{{IBM}}.} \bibinfo{year}{2022}\natexlab{}.
\newblock \bibinfo{title}{{IBM Documentation}}.
\newblock
\newblock
\urldef\tempurl%
\url{https://www.ibm.com/docs/en/spectrum-lsf}
\showURL{%
\tempurl}
\newblock
\shownote{[Online; accessed 14. Oct. 2022]}.


\bibitem[Keplet(2020)]%
        {www-kepler}
\bibfield{author}{\bibinfo{person}{Keplet}.} \bibinfo{year}{2020}\natexlab{}.
\newblock \bibinfo{title}{{The Kepler Project {\ifmmode---\else\textemdash\fi}
  Kepler}}.
\newblock
\newblock
\urldef\tempurl%
\url{https://kepler-project.org/index.html}
\showURL{%
\tempurl}
\newblock
\shownote{[Online; accessed 14. Oct. 2022]}.


\bibitem[Laszewski et~al\mbox{.}(2021)]%
        {las21-gas}
\bibfield{author}{\bibinfo{person}{Gregor~von Laszewski},
  \bibinfo{person}{Anthony Orlowski}, \bibinfo{person}{Richard~H. Otten},
  \bibinfo{person}{Reilly Markowitz}, \bibinfo{person}{Sunny Gandhi},
  \bibinfo{person}{Adam Chai}, \bibinfo{person}{Geoffrey~C. Fox}, {and}
  \bibinfo{person}{Wo~L. Chang}.} \bibinfo{year}{2021}\natexlab{}.
\newblock \showarticletitle{Using Cloudmesh GAS for Speedy Generation of Hybrid
  Multi-Cloud Auto Generated AI Services}. In \bibinfo{booktitle}{\emph{2021
  IEEE 45th Annual Computers, Software, and Applications Conference
  (COMPSAC)}}. \bibinfo{pages}{144--155}.
\newblock
\urldef\tempurl%
\url{https://doi.org/10.1109/COMPSAC51774.2021.00032}
\showDOI{\tempurl}


\bibitem[Merzky et~al\mbox{.}(2015)]%
        {arxiv-radical-pilot}
\bibfield{author}{\bibinfo{person}{Andr{\'{e}} Merzky}, \bibinfo{person}{Mark
  Santcroos}, \bibinfo{person}{Matteo Turilli}, {and} \bibinfo{person}{Shantenu
  Jha}.} \bibinfo{year}{2015}\natexlab{}.
\newblock \showarticletitle{RADICAL-Pilot: Scalable Execution of Heterogeneous
  and Dynamic Workloads on Supercomputers}.
\newblock \bibinfo{journal}{\emph{CoRR}}  \bibinfo{volume}{abs/1512.08194}
  (\bibinfo{year}{2015}).
\newblock
\showeprint[arXiv]{1512.08194}
\urldef\tempurl%
\url{http://arxiv.org/abs/1512.08194}
\showURL{%
\tempurl}


\bibitem[{Microsoft}(2022)]%
        {www-azure-batch}
\bibfield{author}{\bibinfo{person}{{Microsoft}}.}
  \bibinfo{year}{2022}\natexlab{}.
\newblock \bibinfo{title}{{Batch service workflow and resources - Azure
  Batch}}.
\newblock
\newblock
\urldef\tempurl%
\url{https://learn.microsoft.com/en-us/azure/batch/batch-service-workflow-features}
\showURL{%
\tempurl}
\newblock
\shownote{[Online; accessed 14. Oct. 2022]}.


\bibitem[mnist-programs(2022)]%
        {www-mnist-programs}
mnist-programs \bibinfo{year}{2022}\natexlab{}.
\newblock \bibinfo{title}{{UVA C4GC REU 2022 MNIST Programs}}.
\newblock \bibinfo{howpublished}{GitHub}.
\newblock
\urldef\tempurl%
\url{https://github.com/cybertraining-dsc/reu2022/tree/main/code/deeplearning/mnist}
\showURL{%
\tempurl}
\newblock
\shownote{[Online; accessed 24. Oct. 2022]}.


\bibitem[{OASIS}(2022)]%
        {www-wsrf}
\bibfield{author}{\bibinfo{person}{{OASIS}}.} \bibinfo{year}{2022}\natexlab{}.
\newblock \bibinfo{title}{Web Services Resource Framework (WSRF) TC XML Schema
  Documentation}.
\newblock \bibinfo{howpublished}{Web Page}.
\newblock
\urldef\tempurl%
\url{https://schemas.liquid-technologies.com/OASIS/WSRF/1.2/}
\showURL{%
\tempurl}


\bibitem[Outlaw(2016)]%
        {www-azure-rest}
\bibfield{author}{\bibinfo{person}{Robert Outlaw}.}
  \bibinfo{year}{2016}\natexlab{}.
\newblock \bibinfo{title}{{Workflows - REST API (Azure Logic Apps)}}.
\newblock
\newblock
\urldef\tempurl%
\url{https://learn.microsoft.com/en-us/rest/api/logic/workflows}
\showURL{%
\tempurl}
\newblock
\shownote{[Online; accessed 14. Oct. 2022]}.


\bibitem[Papay(2022)]%
        {www-cloudmask}
\bibfield{author}{\bibinfo{person}{Juri Papay}.}
  \bibinfo{year}{2022}\natexlab{}.
\newblock \bibinfo{title}{{CloudMask benchmark notes}}.
\newblock \bibinfo{howpublished}{GitHub}.
\newblock
\urldef\tempurl%
\url{https://github.com/laszewsk/mlcommons/tree/main/benchmarks/cloudmask/#readme}
\showURL{%
\tempurl}
\newblock
\shownote{[Online; accessed 24. Oct. 2022]}.


\bibitem[Pegasus(2022)]%
        {www-pegasus}
Pegasus \bibinfo{year}{2022}\natexlab{}.
\newblock \bibinfo{title}{{Pegasus WMS{'}s documentation}}.
\newblock
\newblock
\urldef\tempurl%
\url{https://pegasus.isi.edu/documentation/index.html}
\showURL{%
\tempurl}
\newblock
\shownote{[Online; accessed 14. Oct. 2022]}.


\bibitem[SLURM(2022)]%
        {www-slurm}
SLURM \bibinfo{year}{2022}\natexlab{}.
\newblock \bibinfo{title}{{Slurm Workload Manager - Documentation}}.
\newblock
\newblock
\urldef\tempurl%
\url{https://slurm.schedmd.com/documentation.html}
\showURL{%
\tempurl}
\newblock
\shownote{[Online; accessed 14. Oct. 2022]}.


\bibitem[snakemake(2022)]%
        {www-snakemake}
snakemake \bibinfo{year}{2022}\natexlab{}.
\newblock \bibinfo{title}{{Snakemake Documentation}}.
\newblock
\newblock
\urldef\tempurl%
\url{https://snakemake.readthedocs.io/en/stable}
\showURL{%
\tempurl}
\newblock
\shownote{[Online; accessed 14. Oct. 2022]}.


\bibitem[von Laszewski(1996)]%
        {las-1996-thesis}
\bibfield{author}{\bibinfo{person}{Gregor von Laszewski}.}
  \bibinfo{year}{1996}\natexlab{}.
\newblock \emph{\bibinfo{title}{A Parallel Data Assimilation System and Its
  Implications on a Metacomputing Environment}}.
\newblock \bibinfo{thesistype}{Ph.\,D. Dissertation}. \bibinfo{school}{Syracuse
  University}, \bibinfo{address}{USA}.
\newblock
\showISBNx{0591478021}
\urldef\tempurl%
\url{https://laszewski.github.io/papers/laszewskithesis.pdf}
\showURL{%
\tempurl}
\newblock
\shownote{AAI9737816}.


\bibitem[von Laszewski(1999)]%
        {las-1999-loosely}
\bibfield{author}{\bibinfo{person}{Gregor von Laszewski}.}
  \bibinfo{year}{1999}\natexlab{}.
\newblock \showarticletitle{A Loosely Coupled Metacomputer: Cooperating Job
  Submissions Across Multiple Supercomputing Sites}.
\newblock \bibinfo{journal}{\emph{Concurrency: Practice and Experience}}
  \bibinfo{volume}{11}, \bibinfo{number}{15} (\bibinfo{date}{December}
  \bibinfo{year}{1999}), \bibinfo{pages}{933--948}.
\newblock
\showISSN{1096-9128}
\urldef\tempurl%
\url{https://doi.org/10.1002/(SICI)1096-9128(19991225)11:15<933::AID-CPE461>3.0.CO;2-J}
\showDOI{\tempurl}
\newblock
\shownote{The initial version of this paper was available in 1996}.


\bibitem[von Laszewski(2019)]%
        {www-cloudmesh-manual}
\bibfield{author}{\bibinfo{person}{Gregor von Laszewski}.}
  \bibinfo{year}{2019}\natexlab{}.
\newblock \bibinfo{title}{Cloudmesh Manual}.
\newblock \bibinfo{howpublished}{GitHub}.
\newblock
\urldef\tempurl%
\url{https://cloudmesh.github.io/cloudmesh-manual/}
\showURL{%
\tempurl}


\bibitem[von Laszewski and Fleischer(2022)]%
        {github-cloudmesh-cc}
\bibfield{author}{\bibinfo{person}{Gregor von Laszewski} {and}
  \bibinfo{person}{J.P. Fleischer}.} \bibinfo{year}{2022}\natexlab{}.
\newblock \bibinfo{title}{{Hybrid Multi-Cloud Analytics Services Framework}}.
\newblock \bibinfo{howpublished}{GitHub}.
\newblock
\urldef\tempurl%
\url{https://github.com/cloudmesh/cloudmesh-cc}
\showURL{%
\tempurl}
\newblock
\shownote{[Online; accessed 14. Oct. 2022]}.


\bibitem[von Laszewski et~al\mbox{.}(2002)]%
        {las-02-infogram}
\bibfield{author}{\bibinfo{person}{Gregor von Laszewski},
  \bibinfo{person}{Jarek Gawor}, \bibinfo{person}{Carlos~J. Peña}, {and}
  \bibinfo{person}{Ian Foster}.} \bibinfo{year}{2002}\natexlab{}.
\newblock \showarticletitle{InfoGram: A Peer-to-Peer Information and Job
  Submission Service}. In \bibinfo{booktitle}{\emph{Proceedings of the 11th
  Symposium on High Performance Distributed Computing}}
  \emph{(\bibinfo{series}{HPDC '02})}. \bibinfo{publisher}{IEEE Computer
  Society and Washington and DC and USA}, \bibinfo{address}{Edinbrough and
  U.K.}, \bibinfo{pages}{333--342}.
\newblock
\showISBNx{0-7695-1686-6}
\urldef\tempurl%
\url{https://laszewski.github.io/papers/vonLaszewski-infogram.pdf}
\showURL{%
\tempurl}
\newblock
\shownote{http://dl.acm.org/citation.cfm?id=822086.823347}.


\bibitem[von Laszewski et~al\mbox{.}(2007)]%
        {las07-workflow}
\bibfield{author}{\bibinfo{person}{Gregor von Laszewski},
  \bibinfo{person}{Mihael Hategan}, {and} \bibinfo{person}{Deepti Kodeboyina}.}
  \bibinfo{year}{2007}\natexlab{}.
\newblock \bibinfo{booktitle}{\emph{Java CoG Kit Workflow}}.
\newblock \bibinfo{publisher}{Springer London}, \bibinfo{address}{London},
  \bibinfo{pages}{340--356}.
\newblock
\showISBNx{978-1-84628-757-2}
\urldef\tempurl%
\url{https://doi.org/10.1007/978-1-84628-757-2_21}
\showDOI{\tempurl}


\bibitem[Zhao et~al\mbox{.}(2007)]%
        {las-2007-swift}
\bibfield{author}{\bibinfo{person}{Yong Zhao}, \bibinfo{person}{Hategan},
  \bibinfo{person}{M.}, \bibinfo{person}{Clifford}, \bibinfo{person}{B.},
  \bibinfo{person}{Foster}, \bibinfo{person}{I.}, \bibinfo{person}{von
  Laszewski}, \bibinfo{person}{G.}, \bibinfo{person}{Nefedova},
  \bibinfo{person}{V.}, \bibinfo{person}{Raicu}, \bibinfo{person}{I.},
  \bibinfo{person}{Stef-Praun}, \bibinfo{person}{T.}, \bibinfo{person}{Wilde},
  {and} \bibinfo{person}{M.}} \bibinfo{year}{2007}\natexlab{}.
\newblock \showarticletitle{Swift: Fast, Reliable and Loosely Coupled Parallel
  Computation}. In \bibinfo{booktitle}{\emph{2007 IEEE Congress on Services}}.
  \bibinfo{pages}{199--206}.
\newblock
\urldef\tempurl%
\url{https://doi.org/10.1109/SERVICES.2007.63}
\showDOI{\tempurl}


\end{thebibliography}

\appendix

\section{Appendix}

\FILE{installation.tex}

\subsection{Installation}\label{installation}

To leverage cloudmesh-cc, use the cloudmesh-installer to install the
Cloudmesh suite of repositories. Optionally, to utilize the graph
visualization you must install also {\em graphviz} must be
installed. On Windows, Git Bash is required, in addition. The overall
installation is very simple and is supported on a variety of operating
systems. We leverage the cloudmesh-installer to locally install the
cloudmesh suite of repositories by executing the following commands:

{\scriptsize\begin{verbatim}
  $ mkdir ~/cm
  $ cd ~/cm
  $ pip install cloudmesh-installer -U
  $ cloudmesh-installer get cc
\end{verbatim}}

To install graphviz you can use the following commands on the
appropriate operating system:

\begin{description}

\item[Windows.]  Git Bash and Graphviz must be installed. The user
can use an instance of Chocolatey that is run as an administrator for
convenience:

{\scriptsize\begin{verbatim}
  $ choco install git.install
          --params "/GitAndUnixToolsOnPath
                    /Editor:Nano
                    /PseudoConsoleSupport
                    /NoAutoCrlf" -y
  $ choco install graphviz -y
\end{verbatim}}

\item[macOS.] Graphviz must be installed. The user can use Homebrew
for convenience:

{\scriptsize\begin{verbatim}
  $ brew install graphviz
\end{verbatim}}

\item[Linux.] Graphviz must be installed. The user can use apt for
convenience:

{\scriptsize\begin{verbatim}
  $ sudo apt install graphviz -y
\end{verbatim}}

\end{description}

To test the workflow program, prepare a {\scriptsize \verb|cm|}
directory in your home directory by executing the following commands:

{\scriptsize\begin{verbatim}
  $ cd ~/cm/cloudmesh-cc
  $ pytest -v -x --capture=no tests
\end{verbatim}}

A variety of separate tests are available that test individual
capabilities.

\FILE{cloudmask-appendix.tex}

\subsection{Running Cloudmask Workflow}\label{sec:running-cloudmask}

To run the Cloudmask workflow, run the following commands:

{\scriptsize\begin{verbatim}
  $ cd ~/cm
  $ git clone https://github.com/laszewsk/mlcommons.git
  $ cd mlcommons
  $ pytest -v -x --capture=no
           benchmarks/cloudmask/target/rivanna/run_cloudmask_workflow.py
\end{verbatim}}

\FILE{contributing.tex}

\subsection{Contributing}\label{contributing}

All contributions are done under the Apache License. The code is
maintained as an open-source project on GitHub while using the typical
GitHub code management tools such as:

\begin{itemize}
\item
  \href{https://github.com/cloudmesh/cloudmesh-cc}{Code Repository}
\item
  \href{https://github.com/cloudmesh/cloudmesh-cc/issues}{Issue Management}
\item
  \href{https://github.com/cloudmesh/cloudmesh-cc/pulls}{Pull Request Management}
\item
  \href{https://github.com/cloudmesh/cloudmesh-cc/actions}{Automatic verification with GitHub Actions}
\end{itemize}

The main branch is the release branch and is supposed to be functional
at all times. Hence, contributions are first done in other branches,
and once agreeing that they need to be integrated into the code, they
are merged into main. All new code must be documented and have
sufficient automated tests. Before creating a pull request, it is
important that the tests within the test directory are passing. The
repository already contains several pytests that can be leveraged to
conduct routine testing of the code, including its SSH remote
functionality, REST capability, and Python interface, among others.

\end{document}